\documentclass[letterpaper,11pt]{article}
\usepackage[margin=1.0in]{geometry}
\usepackage{natbib}
\usepackage{epsfig}
\usepackage{multirow}
\usepackage{graphicx}
\usepackage{array}
\usepackage{color}
\usepackage[table]{xcolor}
\usepackage{rotating}

\newcommand{\specialcell}[2][b]{\begin{tabular}[#1]{@{}c@{}}#2\end{tabular}}

\title{Retrotransposon mobilization in cancer genomes}
\author{Tracy Ballinger\textsuperscript{1}, Adam D. Ewing\textsuperscript{1}, David Haussler\textsuperscript{1,2}
\\ \small \textsuperscript{1}Center for Biomolecular Science and Engineering, University of California, Santa Cruz, CA 95064, USA.
\\ \small \textsuperscript{2}Howard Hughes Medical Institute, University of California, Santa Cruz, CA 95064, USA.
\\}

\begin{document}

\maketitle

\begin{abstract}

The Cancer Genome Atlas project was initiated by the National Cancer Institute in order to characterize the genomes of hundreds of tumors of various cancer types. While much effort has been put into detecting somatic genomic variation in these data, somatic structural variation induced by the activity of transposable element insertions has not been reported. Transposable elements (TEs) are particularly relevant in cancer in part because of several known cases in which a TE insertion is directly linked to cancer formation and studies linking the epigenetic status of retrotransposons to carcinogenesis and patient outcome. Additionally, evidence for somatic retrotransposition in eukaryotic genomes suggests that some tissues and therefore some cancer types may be disposed to increased retrotransposition. We built upon previous work to develop a highly efficient computational pipeline for the detection of non-reference mobile element insertions from high-throughput paired-end whole genome sequencing data that is capable of detecting breakpoints through a local assembly strategy. Using this, we analyzed 33 whole genome tumor datasets with paired normal samples from TCGA across 3 different cancer types: glioblastoma multiforme (GBM), ovarian serous cystoadenocarcinoma (OV) and colorectal adenocarcinoma (COAD). We detected 72 insertions in colon samples, almost all of them LINE-1 elements, and none in GBM or OV. The amount of somatic retrotransposition varies widely between samples with 61 insertions present in one case. The lack of somatic retrotransposon insertions in GBM and OV samples suggests that TE activity in cancer is restricted to certain cancer types.
\end{abstract}

\section{Introduction}

Retrotransposons are found in all eukaryotic genomes. They are observed as repetitive DNA elements due to their capacity to insert new copies of themselves into the host DNA through a copy and paste process using an RNA intermediate~\citep{Boeke:1985p1962}. They are categorized as either long terminal repeat (LTR) or non-LTR and further into families based on sequence similarity to other elements and by their mechanism of mobilization. The non-LTR retrotransposons that inhabit mammalian genomes are likely to mobilize through a mechanism known as target-primed reverse transcription \citep{Luan:1993p1961}. Numerous retrotransposon copies exist in the human genome, comprising at least 45\% of its DNA \citep{Lander:2001p790} and perhaps over two-thirds when highly sensitive TE detection methods are applied \citep{DeKoning2011}. The most prolific retroelements in the human genome include LINE-1 and Alu sequences, comprising 17\% and 8\% of the assembled reference genome, respectively. During primate evolution, the general pattern of retroelement activity has been for one family of LINE element to be active at a time, suggesting either competition for a host factor or adaptation to evade one \citep{Khan2006}. LINE-1 elements are autonomous retrotransposons encoding two proteins \citep{Scott1987} responsible for both their own mobilization \textit{in cis} and the mobilization of non-autonomous Alu elements \citep{Dewannieux2003}, SVA elements \citep{Hancks2011}, and processed pseudogenes \citep{Esnault2000} \textit{in trans}. The activity of the human-specific LINE element, termed L1HS, was first recognized \textit{in vivo} due to its ability to disrupt exons and cause Mendelian disease \citep{Kazazian:1988p1964}. Since then, transposable elements have been linked to a variety of diseases, including cancer, through insertional mutagenesis of exons and regulatory regions near genes, disrupting gene function or regulation (see Hancks et al. 2012 for review). For example, in one case, an exonic L1 insertion was found in the APC tumor suppressor gene in colon cancer tissue but not the normal tissue of the same patient~\citep{YoshioMiki:1992p517}. Intronic retroelement insertions are known to affect splicing by providing 5' or 3' splice sites or disrupting sequence at the branch point \citep{Belancio2008,Hancks2009,Taniguchi-Ikeda2011}. Recent estimates place the rate of L1 retrotransposition in human genomes at 1 new insertion for every 100 to 150 live births~\citep{Ewing:2010p506,Huang:2010p523}. Since retroelements can clearly have an impact on phenotype and disease, it is important that retrotransposon insertion polymorphisms (RIPs) and mutations be characterized in genomic studies. A plethora of recent studies provide various means to document retrotransposon insertion polymorphisms (RIPs) segregating in human populations \citep{Beck:2010p513, Ewing:2010p506, Hormozdiari:2010p524, Huang:2010p523, Iskow:2010p520, Witherspoon:2010p511, Ewing2011, Stewart:2011p1967}, including one report of 9 somatic retrotransposon insertions across 6 lung tumors \citep{Iskow:2010p520}. 

Cancer progression depends on the accumulation of somatic mutations, and recent evidence suggests that retrotransposition also occurs in some somatic tissues such as neuronal stem cells \citep{Muotri:2005p974,Coufal:2009p973,Baillie:2011p1714}. The observation of somatic retrotransposition in specific tissue types suggests tissue-specific regulation, either through known regulators such as APOBEC3 proteins \citep{Kinomoto2007,Muckenfuss2006,Stenglein2006,Chen2006}, germline piRNAs \citep{Aravin2007}, and DNA methylation \citep{Yoder1997, Bourchis:2004p1976}, or through novel mechanism(s) not yet ascribed to transposable elements. Other lines of evidence for somatic retrotransposition include the aforementioned disease-causing insertions, observations of varying levels of transgene-bourne somatic retrotransposition in transgenic mice \citep{Kano2009}, and somatic R2 insertions in \textit{Drosophila simulans} \citep{Eickbush2011}. In addition to mutagenizing both somatic and germline genomes through new insertions, transposable elements play an important role in shaping gene regulatory networks by providing binding sites for transcription factors, including those highly important for cancer progression such as \textit{TP53} and \textit{SOX2} \citep{Wang:2007p93,Bourque:2008p185,Kuwabara2009,Harris2009}. Furthermore, genome-wide methylation status is often assessed through analysis of CpG islands located in the 5' UTRs of LINE-1 elements, which are typically heavily methylated \citep{Woodcock1997}, contributing to their quiescence in most somatic tissue types. Through this assay, a wide variety of cancers are found to be hypomethylated \citep{Ogino2011}, leading us to speculate that retrotransposition rates may be substantially increased in certain cancer types or samples.

In order to test this hypothesis, we took advantage of whole-genome sequence data available through The Cancer Genome Atlas (TCGA). TCGA is an ongoing multi-institutional effort that will eventually include whole genome sequence data for hundreds of tumors and corresponding normal samples for over 20 different cancer types. Here, we consider transposable element insertions in the genomes of three cancer types: glioblastoma multiforme (GBM) \citep{TCGAResearchNetwork2008}, ovarian serous cystoadenocarcinoma (OV) \citep{TCGAResearchNetwork2011}, and colon/rectal adenocarcinoma (COAD/READ) \citep{Muzny2012}, and present evidence for substantially increased retrotransposition in colorectal adenocarcinoma.

\section{Results}

We developed a computational pipeline (discord-retro, http://github.com/adamewing/discord-retro) to detect non-reference retrotransposon insertions from paired end whole genome sequencing data by using mate-pair information from discordantly mapped read pairs (see Methods). We measured the detection characteristics of our application by repeatedly inserting 100 retrotransposons into the euchromatic portion of human chr22 at random positions, generating paired reads, mapping to the GRCh37 reference sequence, and applying our method (see Methods for details). We observe 87.9\% sensitivity with perfect specificity when insertions into other insertions of the same class (e.g. LINE into a LINE or Alu into an Alu) are discarded, and 94.5\% sensitivity and perfect specificity if these insertions are allowed. Using discord-retro, we analyzed 33 high coverage (\textgreater 30x) tumor and normal genome pairs produced by TCGA, and identified retrotransposon insertions not found in the human reference genome (NCBI36 or GRCh37). For high coverage data, the tumor and patient-matched normal paired-end sequence data were combined in order to distinguish between a non-reference germline insertion, which would be found in both tissues, from a somatic insertion, which would be found only in the tumor (or normal) DNA. Refinement of junctions using local assembly and analysis of soft-clipped reads allowed breakpoint-resolution on one or both ends of a predicted insertion. In 45\% of cases, we identified target site duplications (TSDs), a short duplication of sequence around the insertion site that occur as a byproduct of target primed reverse transcription.

\subsection{Germline Insertions}

Across the 33 tumor/normal pairs, we identified 7022 non-reference retrotransposon insertions present in both the tumor and corresponding normal genome, of which 3273 overlapped with a previous study \citep{Wang:2006p1119, Beck:2010p513, Ewing:2010p506, Hormozdiari:2010p524, Iskow:2010p520, Witherspoon:2010p511, Ewing2011, Stewart:2011p1967} (Fig S1).  Of all insertions detected, 727 were LINE insertions, 6101 were Alu insertions and 189 were SVA insertions. Of the 3749 previously uncataloged insertions, 350 were LINE-1 elements, 3220 were Alu elements, and 177 were SVA elements. For every tumor/normal pair we detected an average of 111 LINEs, 823 Alus, and 36 SVAs (Fig. 1). We detected an average of 8.4 LINE, 79.6 Alu and 1.9 SVA insertions that were present in only one sample. The chromosomal distribution of insertions is illustrated in Figure 2.

\subsection{Somatic Insertions}

Somatic insertions are those occuring exclusively in either the tumor or normal sample of a patient-matched pair of genomes and also not present in any other sample or catalogue of retrotransposon insertions from a previous study. Furthermore, because we combine discordant read pairs across both the tumor and normal tissue for an individual, we can be sure that if a tumor-specific or normal-specific call is made, not a single read that could indicate the presence of the insertion exists in the other sample. We found 72 tumor-specific LINE-1 insertions from 4 colon cancer tumor/normal pairs and none from any of the 18 GBM and 10 OV tumor/normal pairs (Fig. 3). Conversely, for the four samples with tumor-specific insertions, we found very few insertions present only in the normal sample (Fig. S2). For one sample, TCGA-AA-A00R, there was an abnormally high number of normal-specific predictions which we believe to be an artifact. The distribution of insertion lengths for the tumor-specific LINE-1 insertions differs markedly from insertions found in both the tumor and normal tissues (Fig. 4).  Of 655 LINE-1 insertions present in both tumor and normal genomes, 147 were full length, defined here as 5.8 kb of greater as indicated by the minimum and maximum mapping locations of paired reads within the reference elements, which resembles the distribution expected from the length distribution in the reference genome \citep{Grimaldi1984,Pavlicek2002}. In contrast, only 2 out of 72 tumor-specific LINE-1 insertions were full-length ($p < 9.68\times10^{-6}$, Fisher's exact test).

Of the 72 tumor-specific insertions, one occurred in the 3' UTR of \textit{PPP1R1C} (protein phosphatase 1, regulator subunit 1C), and 22 occurred in introns, including an insertion in \textit{NAV3}, a gene associated with colon cancer \citep{Carlsson2012}. The \textit{NAV3} insertion occurs in patient TCGA-AA-3518 115 bp downstream of the third exon in the same orientation as the gene in a region overlapped by DNAse hypersensitivity and H3K27 acetylation signals (Fig. S3), indicating possible regulatory elements nearby. An examination of gene expression levels (Agilent 244K Custom Gene Expression G4502A-07-3) between the tumor and normal insertions was carried out using the UCSC Cancer Genome Browser \citep{Zhu2009}, which indicated lower expression of \textit{NAV3} and other genes containing tumor-specific insertions (notably \textit{A2BP1}, and \textit{CTNNA2}) in the tumor relative to the patient-matched normal colon tissue (Fig. 5). Here we focus further analysis on \textit{NAV3}, as decreases in its expression are frequent in colorectal adenomas \citep{Carlsson2012}. Analysis of Agilent expression data from TCGA-AA-3518 shows the difference in expression between \textit{NAV3} in tumor and normal tissues is ranked at the 93rd percentile relative to differences between all other genes in the same tumor/normal pair. To ascertain whether other somatic mutations might be responsible for the observed change in expression, we compared \textit{NAV3} and the surrounding region between cancer and normal genomes of TCGA-AA-3518 using the BamBam algorithm (Z. Sanborn, unpublished).  We detected evidence for 9 potentially cancer-specific SNPs in the 382 kbp region (Fig. S4), but we found no evidence for point mutations or CNVs in exons or in the proximal upstream region likely to have an obvious effect on transcript abundance apart from the L1 insertion. 

\subsection{Similarity to reference elements}

After acquiring a set of insertion predictions for each sample, we sought to determine the closest element in the reference genome in terms of sequence similarity, as this may represent an element similar to the active progenitor element. In general, it is unlikely that the true progenitor can be identified through sequence similarity alone, as the active elements in the human reference genome diverge from one another by less than 1\% \citep{Brouha2003,Seleme2006,Beck:2010p513}. That said, identification of the most similar elements in the human reference genome based on local re-assembly of the elements detected by discord-retro yields an enrichment of full-length, intact, human L1 elements, some of which are known to be active elements (Table 1). This serves as further evidence to substantiate our claim that our report of 72 tumor-specific L1 insertions in 4 colon cancer cases are novel insertions derived from active L1 elements.

\section{Discussion}

As TCGA and others continue to sequence more cancer and paired normal cases across a wider number of cancer types, we may uncover clear driver mutations caused by transposable elements and other cancer types that exhibit high levels of insertional mutagenesis by transposable elements. It is remarkable that TE activity appears so much higher in colorectal adenocarcinomas than in glioblastoma multiforme or ovarian serous cystoadenocarcinoma, but the specific mechanism behind this tissue specificity has eluded us so far.  An intronic insertion in \textit{NAV3} seemes to be paired with a marked decrease in gene expression, although it is far from clear if there is a direct relationship between the presence of somatic LINE-1 insertions and the expression decrease in cases like this. Given that a survey of other somatic mutations in and surrounding \textit{NAV3} yielded nothing that stood out as a possibly expression-altering mutation, and an insertion in \textit{NAV3} occured only 112 bp downstream of an exon in a region with an epigenetic profile indicating regulatory potential, we posit that the LINE-1 insertion may be responsible for the cancer-specific decrease in expression in this instance.

Our general knowledge of somatic retrotransposition is limited, although new technologies and the decreasing cost of sequencing will likely provide new insights in the near future as sequencing studies begin to focus on multiple tissues from a single donor individual. In most respects, the somatic tumor-specific insertions detected by our method are similar to germline insertions, with the notable exception of their length: 97.2\% of tumor-specific insertions are truncated as compared to a 77.6\% truncation rate for germline insertions (insertions detected in both normal and cancer samples). The mechanism for L1 truncation is unknown; conjectures include the presence of an endo- or exonuclease that targets the L1 RNA template, or a factor that interferes with reverse-transcription. At this stage the etiology of element truncation in colon tumors and how it differs from normal or germline tissue is unknown. It may be illuminating to work out why tumor-specific insertions are more severely truncated than those in the germline as a future study and whether this is a general characterisic of somatic retrotransposition or if there is some connection to tumor biology.

This is an exciting time for transposable element biology given our improving ability to explore entire genomes. In this case, whole-genome paired-end sequencing has allowed us to detect somatic retrotransposition in cancer genomes, an observation that opens many new questions regarding the role of mobile DNA in carcinogenesis and tumor molecular biology. As sequencing technologies and out ability to detect structural variants improve, so will our ability to characterize new TE insertions and their parent elements, perhaps gaining further insight on what leads to tissue or disease specific TE activation.

\section{Methods}

A number of successful computational methods have been devised capable of detecting transposable element insertions from whole-genome sequence data including VariationHunter2 \citep{Hormozdiari:2010p522}, T-lex \citep{Fiston-Lavier2011}, RetroSeq (https://github.com/tk2/RetroSeq), HYDRA-SV \citep{Quinlan:2010p515}, and Tea \citep{Lee2012}. The approach outlined here, implemented as discord-retro, has the advantage of working directly from the ubiquitous .bam sequence alignment format (as does RetroSeq) with minimal need for additional mapping apart from that required to identify insertion breakpoints. Here, we give a high-level overview of our method. Sequence data analyzed in this study was generated on the Illumina platform and aligned to a human reference assembly (NCBI36 or GRCh37) by TCGA Research Network members at TCGA Genome Sequencing Centers.

Paired-end reads can be classified based on how they map to the reference genome. A read pair is called concordant if both reads map the proper distance apart and in the correct orientation for the insert size and procedure used in the library preparation and sequencing, and discordant if these conditions are not met. For example, ends of a discordant paired read may map to different chromosomes, too far apart, too close together, or in the wrong orientation. A second type of improperly paired reads are ones in which one read maps to the reference, but its pair does not. These reads are referred to as one-end-anchored (OEA). Lastly, reads are called soft-clipped when part of the read aligns to the reference sequence, but either or both ends of the read do not. 

We first selected all discordant reads from both the tumor and normal sequencing data of a patient where one read of the pair maps to a unique portion of the genome, called the ``anchored'' read, and the other end maps to a repeatmasker annotation elsewhere in the genome. We will refer to these types of read-pairs as one-end-repeat (OER) reads.  We filtered elements corresponding to AluS and LTR elements from the results due to an overabundance of calls with no corresponding breakpoint predictions in some samples. Regions where the uniquely mapped ends of the OER reads clustered in two peaks with opposite orientation were considered consistent with an insertion existing between the two clusters of OER reads. We require there to be 8 OER read pairs within a 500bp window, and for there to be at least 2 uniquely mapped or ``anchored'' reads on either strand. The requirement that both breakpoints (5' and 3' junctions) be covered by paired reads reduces the chance of incorrectly annotating a segmental duplication, translocation, or inversion as a transposable element insertion.

The selection of clustered discordant OER reads yields a set of ~20-50bp windows as predicted transposable element insertion sites. These were annotated as ``germline'' if there were discordant reads in both the tumor and normal tissue samples, as ``somatic/cancer'' if there were contributing discordant reads only in the tumor tissue, and as ``somatic/normal'' if there were contributing discordant reads only in the normal tissue. Insertion loci are cross-referenced against retrotransposon insertion polymorphisms (RIPs) cataloged in previous studies  \citep{Wang:2006p1119, Beck:2010p513, Ewing:2010p506, Hormozdiari:2010p524, Iskow:2010p520, Witherspoon:2010p511, Ewing2011, Stewart:2011p1967} and against each other. As breakpoint resolution varies across studies, insertions within the same 500bp window were considered overlapping. A total of 14 (16\%) potentially tumor-specific insertions were eliminated by comparison to known RIPs and RIPs found in this study.

\subsection{Breakpoint refinement using soft-clipped reads}

Soft-clipped reads mapped using bwa \citep{Li:2009p1125} could be used to pinpoint a breakpoint in the insertion site. For each of the 14 samples aligned with bwa (Table S1), soft-clipped reads mapping within 500bp of each of the predicted insertion sites and which had greater than 10 bp clipped from the read were used to find a consensus breakpoint where a majority of soft-clipped read ends occurred at the same nucleotide in the reference genome. When breakpoints for both the 5' and 3' junctions between the element and the reference genome were detected, we identified target site duplications when the breakpoint on the forward strand occured 3-50bp downstream of the breakpoint on the reverse strand. 

\subsection{Breakpoint refinement with local assembly}
\label{assembly-methods}
We used a local assembly and realignment strategy to determine breakpoints for all samples. All discordant and soft-clipped reads within 500bp of a predicted insertion site were assembled using Velvet \citep{Zerbino:2008p1191} with a k-mer size of 31, the shortPaired option, and insert length of 300. If the reads assembled into 5 contigs or less, these contigs were mapped back to the reference genome using BLAT.  A cutoff of 5 contigs was chosen because when more contigs were present, they were generally too short to be more informative than the original reads.  After mapping the assembled contigs back to the reference assembly, breakpoints present as the point where a contig no longer matches the reference sequence and begins matching a reference retroelement sequence. Target site duplications could be ascertained in cases where two assembled contigs had overlapping alignments to the predicted insertion site on opposite strands. 

\subsection{Simulation}

To measure the accuracy and sensitivity of our pipeline, we inserted 100 LINE, SINE, and SVA sequences randomly into the euchromatic sequence of chr22 from hg19/GRCh37. The retroelement sequences were randomly truncated on the 5' end up to 75\% of the original element length for LINEs and SVAs, and 25\% for Alus. Poly(A) tails between 20 and 70bp in length were added to the 3' end, and 12bp of the target insertion site was duplicated on the 5' junction to mimic target site duplications. Paired Illumina reads were simulated via wgsim (https://github.com/lh3/wgsim) to generate paired 75bp reads at 30x coverage. These reads were mapped back to the reference genome using bwa~\citep{Li:2009p1125} with the following parameters: \texttt{-q 5 -l 32 -k 2 -t 4 -o 1}, alignments were processed with samtools \citep{Li2009} and used as input to discord-retro.

\subsection{Assessing sequence similarity to reference elements}

We performed local sequence assembly as described in section~\ref{assembly-methods} to generate contigs corresponding to inserted sequences. BLAT alignments of the contigs were carried out to find the most closely related elements in the reference genome. Repeatmasker-annotated elements were scored by the sum of the products of the percent identity of a BLAT alignment and the length of the alignment for each contig that overlapped the repeat masked element. The element with the highest score was predicted to be the source element for the new insertion, excluding elements within 1000bp of the insertion site.  Elements scoring within 20 percent-identity bases of the highest score were considered as potential progenitors as well. In cases where there were several repeat elements tied for the highest score or very close to the highest score, the progenitor is considered ambiguous.  We ranked the repeat masked elements by the number of times they were predicted to be a progenitor for an somatic insertion, whether ambiguous or not, and examined the top 10 elements for retrotransposition capability. 

\bibliographystyle{genres}
\bibliography{retropaper}
\newpage

\section{Figures}

\begin{figure}
\epsfig{file=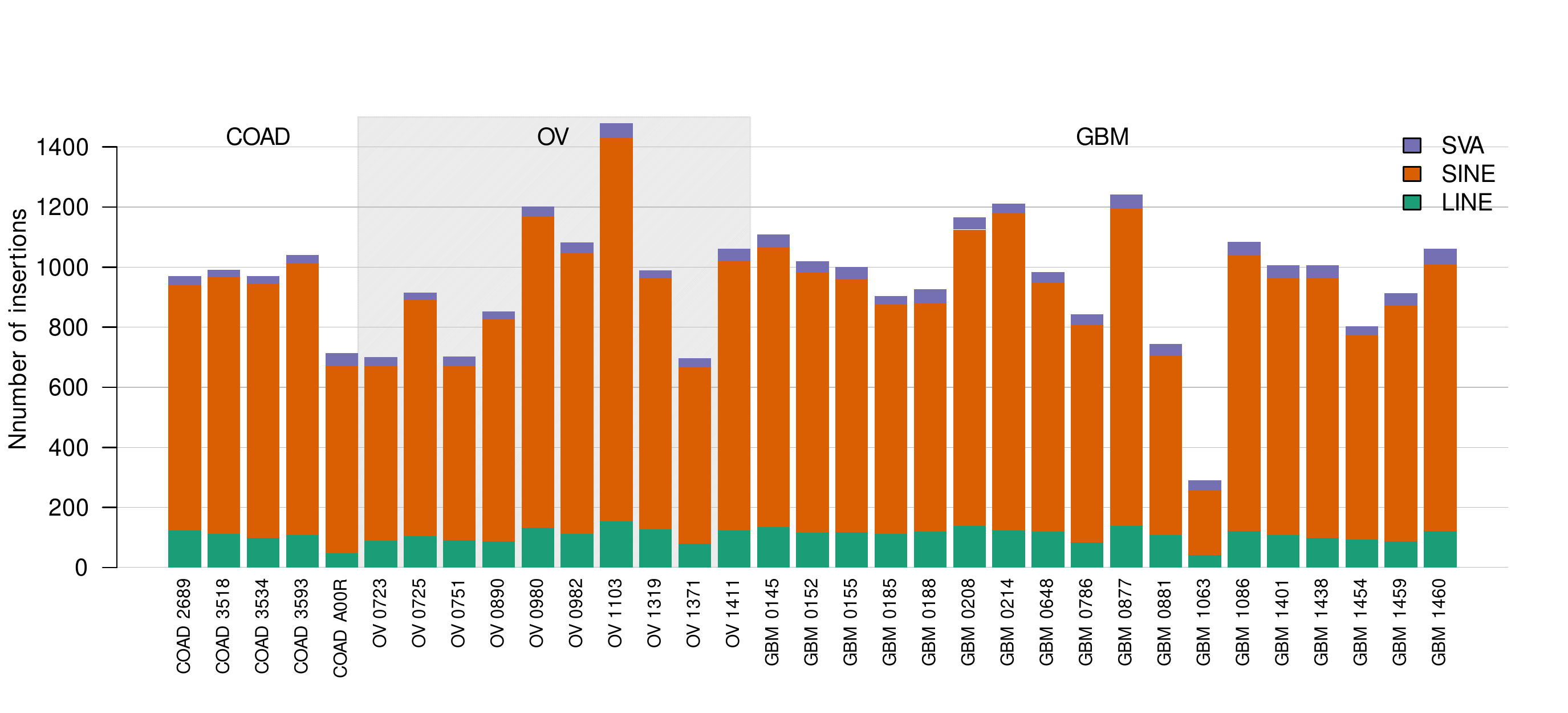, width=18cm}
\caption[]{\label{fig1} The number of non-reference germline insertions (found in both normal and tumor samples) per patient analyzed. }
\end{figure}

\begin{figure}
\epsfig{file=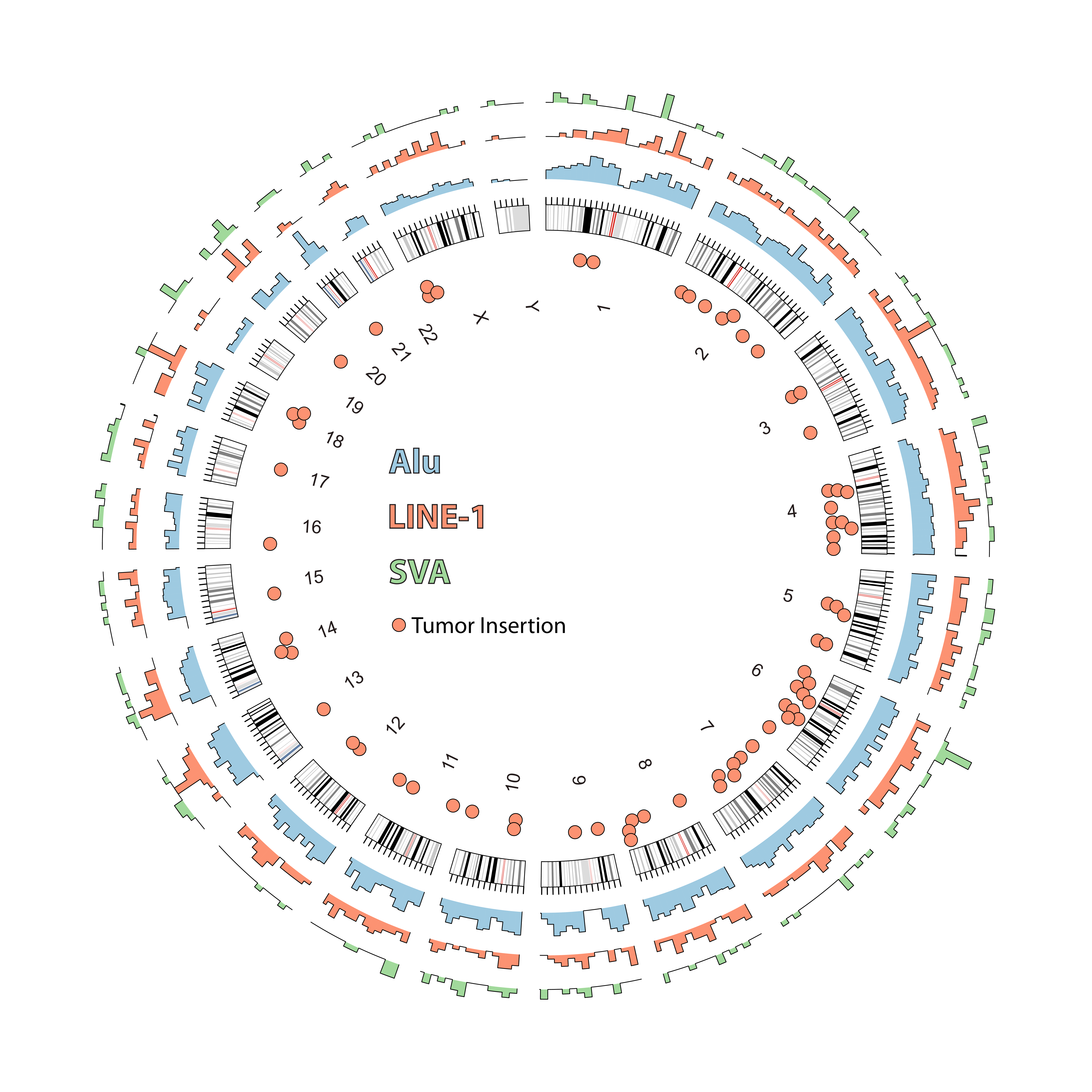, width=18cm}
\caption[]{\label{fig2} Plot depicting distribution of insertion site density for non-reference mobile element insertions (outer rings, colored as indicated) and tumor-specific insertions (inner ring, circles).}
\end{figure}

\begin{figure}
\epsfig{file=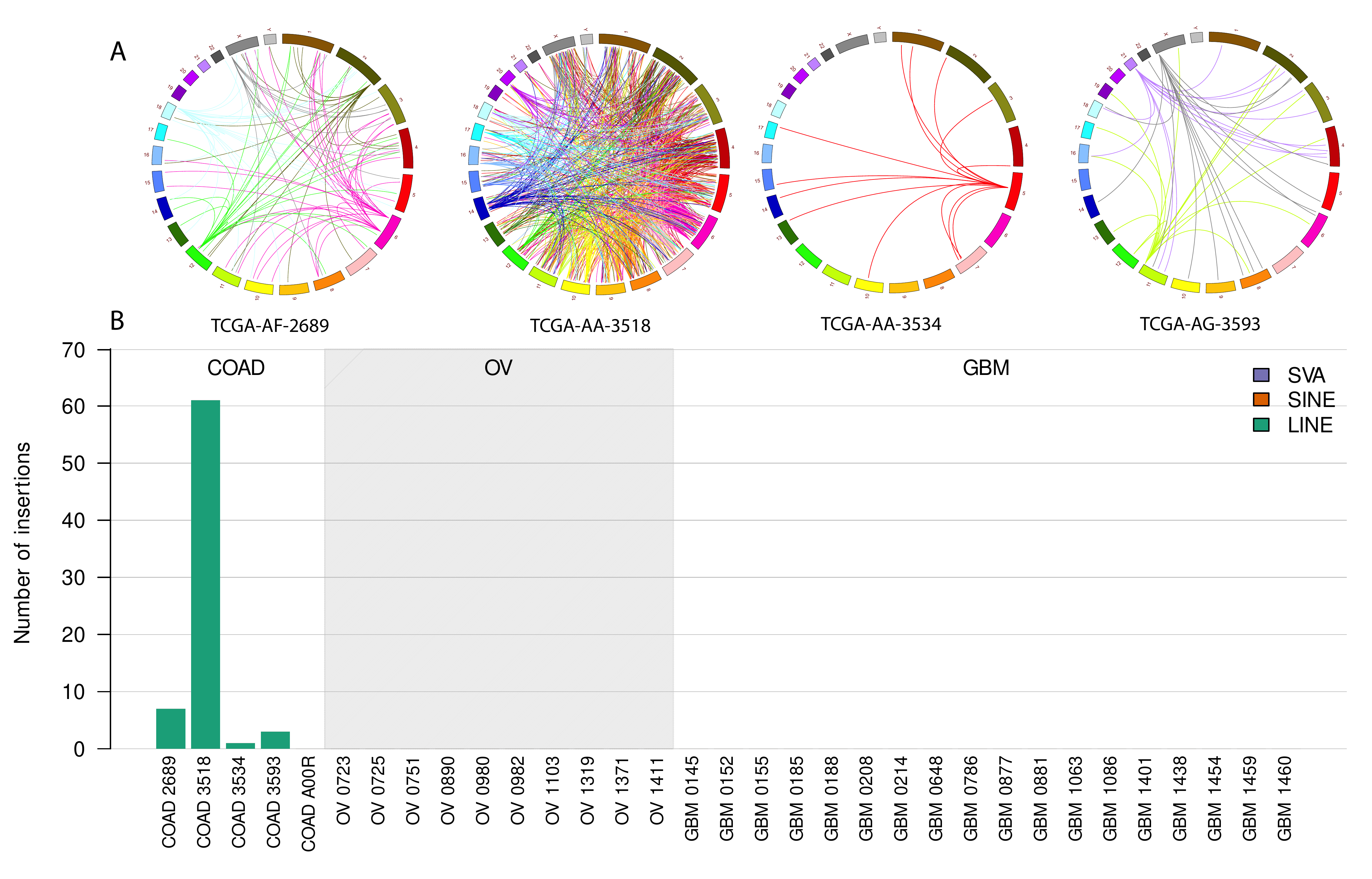, width=18cm}
\caption[]{\label{fig3} (A) Discordant read ``one-end repeat (OER)" mappings for the 4 colorectal adenocarcinoma samples with tumor-specific retrotranposon activity. Links are shown in the color of the chromosome where the insertion occurred.
(B) The number of retrotransposon insertions detected only in the tumor tissue for each of 33 patients analyzed.  No cancer-specific Alu or SVA insertions were detected, and no insertions were found in ovarian carcinoma (OV) or glioblastoma multiforme (GBM), but between 1 and 61 cancer-specific insertions were found across 4 colorectal adenocarcinoma (COAD) patients� tumors.}
\end{figure}

\begin{figure}
\epsfig{file=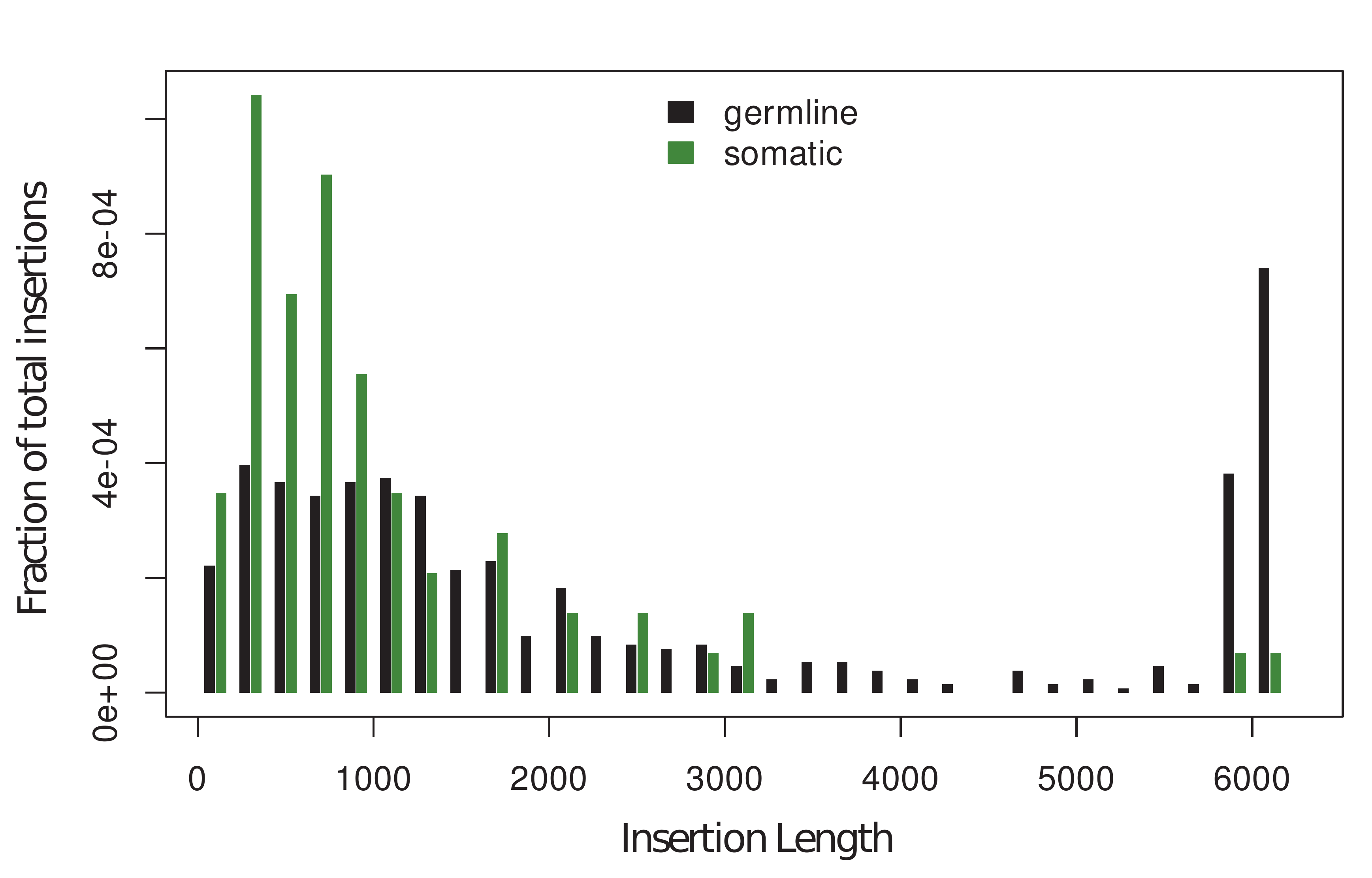, width=18cm}
\caption[]{\label{fig4} The lengths of non-reference L1 insertions detected in this study, binned in 200bp intervals.  Germline insertions are found in the normal and the tumor tissue of a patient, found in a previous study, or found in multiple patients, and somatic/cancer insertions are found only in the tumor tissue of a single patient.  There is a significant difference between the proportion of L1 germline insertions that are full length (\textgreater 5.8kbp) and the proportion of somatic/cancer L1 insertions that are full length (p-value=9.68 x 10-6). }
\end{figure}

\begin{figure}
\epsfig{file=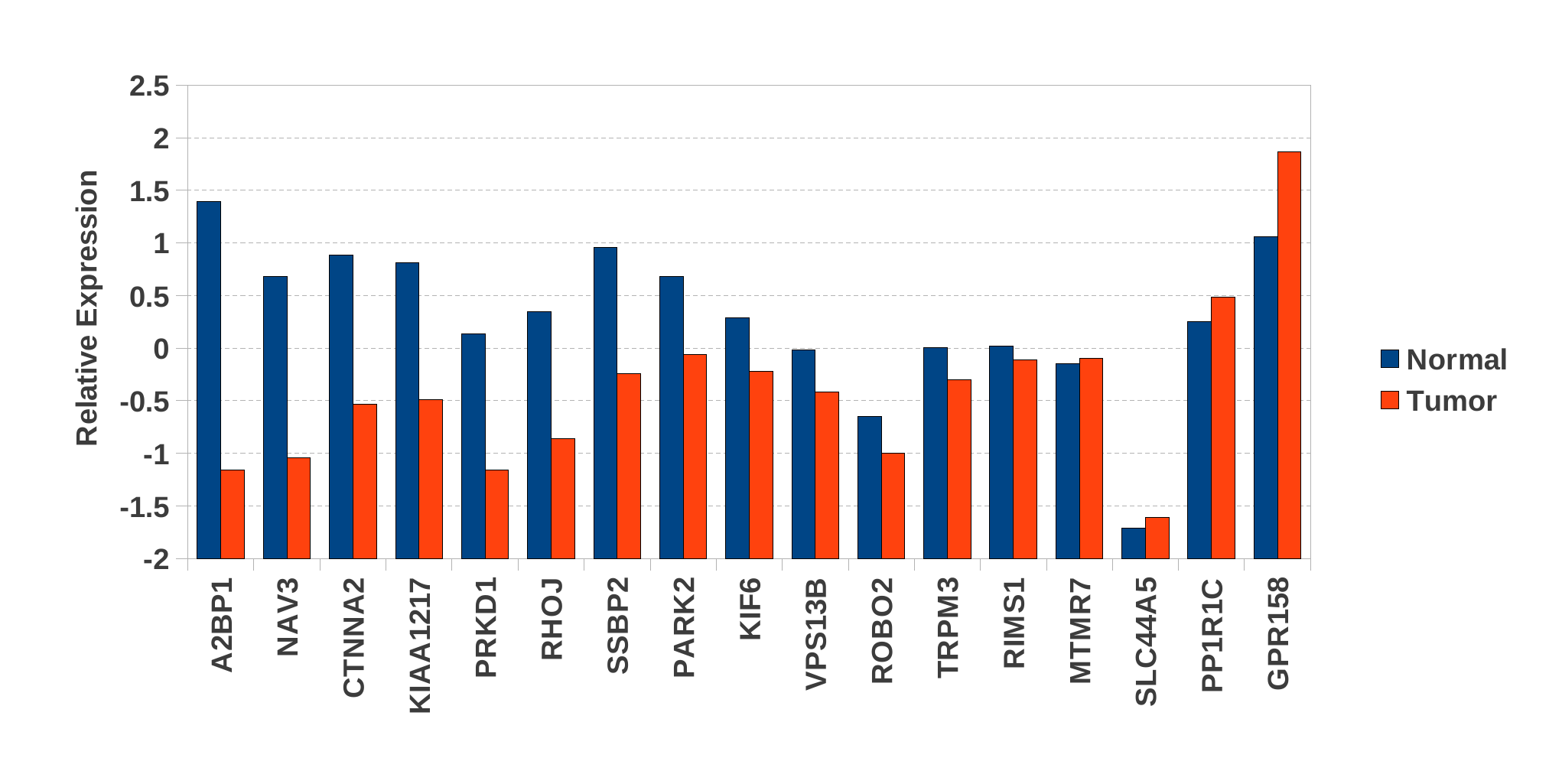, width=18cm}
\caption[]{\label{fig5} Expression level of genes in normal tissue (blue) prior to intronic LINE-1 insertion, and after (red). Relative expression values are taken from the UCSC Cancer Genome Browser (Zhu et al. 2009), where they were normalized by centering to the mean expression level.}
\end{figure}

\section{Tables}

\begin{table}
\noindent\makebox[\textwidth] {% 
\scalebox{1}{%
\begin{tabular}{l|lccccl}
\multicolumn{7}{c}{LINE elements most closely related to inserted sequences} \\
\hline
 & Location & \specialcell{Repeat\\Family} & Length & \specialcell{Best Related\\Insertions} & \specialcell{Related\\Insertions} & Characteristics \\
\hline
\hline
\baselineskip=12pt
 \cellcolor[gray]{0.9}1 & \baselineskip=12pt
 \cellcolor[gray]{0.9}chr10:107127095-107133125 & \baselineskip=12pt
 \cellcolor[gray]{0.9}L1HS & \baselineskip=12pt
 \cellcolor[gray]{0.9}6030 & \baselineskip=12pt
 \cellcolor[gray]{0.9}14 & \baselineskip=12pt
 \cellcolor[gray]{0.9}24 & \baselineskip=12pt
 \cellcolor[gray]{0.9}ORF1 broken, ORF2 intact \\
\baselineskip=12pt
 2 & \baselineskip=12pt
 chr11:92793801-92799845 & \baselineskip=12pt
 L1HS & \baselineskip=12pt
 6044 & \baselineskip=12pt
 12 & \baselineskip=12pt
 33 & \baselineskip=12pt
 intact \\
\baselineskip=12pt
 \cellcolor[gray]{0.9}3 & \baselineskip=12pt
 \cellcolor[gray]{0.9}chr11:24306074-24312123 & \baselineskip=12pt
 \cellcolor[gray]{0.9}L1HS & \baselineskip=12pt
 \cellcolor[gray]{0.9}6049 & \baselineskip=12pt
 \cellcolor[gray]{0.9}12 & \baselineskip=12pt
 \cellcolor[gray]{0.9}28 & \baselineskip=12pt
 \cellcolor[gray]{0.9}intact \\
\baselineskip=12pt
 4 & \baselineskip=12pt
 chr17:65966693-65972723 & \baselineskip=12pt
 L1HS & \baselineskip=12pt
 6030 & \baselineskip=12pt
 10 & \baselineskip=12pt
 26 & \baselineskip=12pt
 ORF1 intact, ORF2 broken \\
\baselineskip=12pt
 \cellcolor[gray]{0.9}5 & \baselineskip=12pt
 \cellcolor[gray]{0.9}chr11:60608423-60610418 & \baselineskip=12pt
 \cellcolor[gray]{0.9}L1HS & \baselineskip=12pt
 \cellcolor[gray]{0.9}1995 & \baselineskip=12pt
 \cellcolor[gray]{0.9}10 & \baselineskip=12pt
 \cellcolor[gray]{0.9}18 & \baselineskip=12pt
 \cellcolor[gray]{0.9}truncated \\
\baselineskip=12pt
 6 & \baselineskip=12pt
 chr7:49690411-49696442 & \baselineskip=12pt
 L1HS & \baselineskip=12pt
 6031 & \baselineskip=12pt
 10 & \baselineskip=12pt
 18 & \baselineskip=12pt
 intact \\
\baselineskip=12pt
 \cellcolor[gray]{0.9}7 & \baselineskip=12pt
 \cellcolor[gray]{0.9}chr16:22618776-22619548 & \baselineskip=12pt
 \cellcolor[gray]{0.9}L1HS & \baselineskip=12pt
 \cellcolor[gray]{0.9}772 & \baselineskip=12pt
 \cellcolor[gray]{0.9}8 & \baselineskip=12pt
 \cellcolor[gray]{0.9}14 & \baselineskip=12pt
 \cellcolor[gray]{0.9}truncated \\
\baselineskip=12pt
 8 & \baselineskip=12pt
 chr18:43440743-43446771 & \baselineskip=12pt
 L1HS & \baselineskip=12pt
 6028 & \baselineskip=12pt
 7 & \baselineskip=12pt
 19 & \baselineskip=12pt
 ORF1 intact, ORF2 broken \\
\baselineskip=12pt
 \cellcolor[gray]{0.9}9 & \baselineskip=12pt
 \cellcolor[gray]{0.9}chr4:182113842-182114873 & \baselineskip=12pt
 \cellcolor[gray]{0.9}L1HS & \baselineskip=12pt
 \cellcolor[gray]{0.9}1031 & \baselineskip=12pt
 \cellcolor[gray]{0.9}7 & \baselineskip=12pt
 \cellcolor[gray]{0.9}11 & \baselineskip=12pt
 \cellcolor[gray]{0.9}truncated \\
\baselineskip=12pt
 10 & \baselineskip=12pt
 chr5:12250459-12251104 & \baselineskip=12pt
 L1HS & \baselineskip=12pt
 645 & \baselineskip=12pt
 7 & \baselineskip=12pt
 9 & \baselineskip=12pt
 truncated \\
\hline

\end{tabular}
}
}
\caption{Contigs of inserted sequence were assembled for some cancer-specific transposable element (TE) insertions and aligned back to the reference genome using BLAT.  Repeat masker annotated elements from the reference genome are listed according to the number of times they have the highest sequence similarity to an insertion�s contigs compared to all other repeat masker annotated elements (Best Related Insertions).  The number of times an element has within 20 mismatches of the highest sequence similarity to an insertion�s contigs (in essence, is the most similar or a close second), is also listed (Related Insertions).}
\label{Reference elements with the most sequence similarity to tumor-specific insertions}
\end{table}

\end{document}